\documentclass[doublespacing,review]{elsarticle}
\usepackage{amssymb}
\usepackage{amsmath}
\usepackage{amsthm}
\usepackage{graphics}

\journal{Physics Letters A}

\begin{document}

\begin{frontmatter}



\title{Effect of finite Coulomb interaction on full counting statistics of electronic
transport through single-molecule magnet}

\author{Hai-Bin Xue}
\ead {xhb98326110@163.com}
\author{Y.-H. Nie\corref{cor1}}
\cortext[cor1]{Corresponding author. Tel.: +86 351 7011399; fax: +86
351 7011399.}
\ead {nieyh@sxu.edu.cn}
\author{Z.-J. Li}
\author{J.-Q. Liang}
\address{Institute of Theoretical Physics, Shanxi University, Taiyuan, Shanxi 030006, China}

\begin{abstract}

We study the full counting statistics (FCS) in a single-molecule
magnet (SMM) with finite Coulomb interaction $U$. For finite $U$ the
FCS, differing from $U\rightarrow \infty $, shows a symmetric
gate-voltage-dependence when the coupling strengths with two
electrodes are interchanged, which can be observed experimentally
just by reversing the bias-voltage. Moreover, we find that the
effect of finite $U$ on shot noise depends on the internal level
structure of the SMM and the coupling asymmetry of the SMM with two
electrodes as well. When the coupling of the SMM with the
incident-electrode is stronger than that with the
outgoing-electrode, the super-Poissonian shot noise in the
sequential tunneling regime appears under relatively small
gate-voltage and relatively large finite $U$, and dose not for
$U\rightarrow \infty $; while it occurs at relatively large
gate-voltage for the opposite coupling case. The formation mechanism
of super-Poissonian shot noise can be qualitatively attributed to
the competition between fast and slow transport channels.

\end{abstract}


\end{frontmatter}

\newpage

\section{Introduction}

Electronic transport through a single-molecule magnet (SMM) has been
intensively studied both experimentally\cite{Heersche} and theoretically\cite%
{Romeike1,Romeike2,Elste,Timm,Xue,Leuenberger,Misiorny,LuHZ}
stimulated by the prospect of a new generation of molecule-based
electronic and spintronic devices\cite{Lapo}. Recently, the current
fluctuation in electron transport through single-molecule magnet or
molecular junction has been attracting much
interest\cite{Romeike2,Xue,Djukic,Imura,Welack,Dong,Aguado} owing to
its allowing
one to identify the internal level structure of the transport system\cite%
{Belzig} and to access information of electron correlation that can
not be
contained in the differential conductance and the average current\cite%
{Blanter}. These studies were mainly focused on the infinite Coulomb
interaction regime\cite{Timm,Xue}. In fact, Coulomb interaction is
usually finite in a realistic mesoscopic system and thus one should
consider the effects of the finite Coulomb interaction on the
current correlation. In most mesoscopic systems, the negative
correlation induced by Coulomb interaction may impose a time delay
between two consecutive electron transfers and lead to the
suppression of shot noise so that sub-Poissonian
shot noise occurs. For example, in symmetrical double-barrier junctions\cite%
{Chen} and in nondegenerate diffusive conductors\cite{Gonzalez} the
1/2 and 1/3 suppression Fano-factors have been found, respectively.
However, in the presence of a strong nonlinearity of the $I$-$V$
characteristics, the Coulomb interaction can yield a positive
correlation and enhance the noise even to be
super-Poissonian\cite{Iannaccone,Reklaitis}. This phenomenon was
first discovered in double-barrier tunneling diodes in the negative
differential conductance (NDC) regime and a Fano factor up to 6.6
was observed\cite{Iannaccone}. In the NDC regime, Coulomb
interaction and the density-shape of states in the well introduce
positive correlation between consecutive current pulses, which leads
to a super-Poissonian shot noise. In general, the effect of Coulomb
interaction on the shot noise is more complicated. For instance, in
mesoscopic coherent conductors (at sufficiently large voltages)
Coulomb interaction decreases the shot noise at low transmissions
and increases it at high transmissions\cite{Galaktionov}. Shot-noise
measurements in mesoscopic devices can also provide information
about the effective charge $e^{\ast }$ of the current-carrying
particles. For a quantum-dot system in the Kondo regime, the
simultaneous presence of
one- and two-quasiparticle scattering results in a universal average charge $%
e^{\ast }=5/3e$\cite{Sela}. However, as shown in Ref. \cite{Golub},
the Coulomb interaction remarkably influences the effective
backscattering charge of current-carrying particles via a correction
factor ($e^{\ast }=5/3eF$) which is less than unity. Furthermore,
the effect of Coulomb interaction on the shot noise can be employed
to reveal important information of the energy profile of
nonequilibrium carriers injected from an emitter contact, but which
can not be obtained from shot-noise measurements in the absence of
Coulomb interactions\cite{Naspreda}. In the present SMM system with
finite Coulomb energy $U$, an electron from one lead tunnels into
the SMM and then leaves the SMM through the other lead via two kinds
of transition processes: (i) between the singly-occupied and empty
states, (ii) between doubly-occupied and the singly-occupied states
which does not occur under the infinite Coulomb interaction.
Therefore, it is significant to study the effect of finite Coulomb
interaction on the current fluctuation in the SMM system.

An alternative way to investigate the current fluctuation, known as
the full counting statistics (FCS), has been proposed in the
pioneering work by Levitov \emph{et al.}\cite{Levitov}. The method
yields not only shot-noise power but also all the statistical
cumulants of the number of transferred charges. The transport
through mesoscopic devices is fully described by the FCS, which may
provide the full information about the probability distribution
$P\left( n,t\right) $ of transferring $n$ electrons between
electrode and SMM during a time interval $t$. The FCS is obtained
from the cumulant generating function (CGF) that is related to the
probability
distribution by\cite{Bagrets}%
\begin{equation}
e^{-F\left( \chi\right) }=\sum_{n}P\left( n,t\right) e^{in\chi },
\label{CGF}
\end{equation}
where $\chi$ is the counting field. All cumulants of the current can
be obtained from the CGF by performing derivatives with respect to
the counting field $C_{k}=\left. -\left( -i\partial_{\chi}\right)
^{k}F\left( \chi\right) \right\vert _{\chi=0}$. In the long-time
limit, the first three cumulants are directly related to the
transport characteristics. For example, the first-order cumulant
(the peak position of the distribution of
transferred-electron number) $C_{1}=\bar{n}$ gives the average current $%
\left\langle I\right\rangle =eC_{1}/t$. The zero-frequency shot
noise is
related to the second-order cumulant (the peak-width of the distribution) $%
S=2e^{2}C_{2}/t=2e^{2}\left( \overline{n^{2}}-\bar{n}^{2}\right)
/t$. The third cumulant $C_{3}=\overline{\left( n-\bar{n}\right)
^{3}}$ characterizes
the skewness of the distribution. Here, $\overline{\left( \cdots\right) }%
=\sum_{n}\left( \cdots\right) P\left( n,t\right) $. In general, the
shot
noise and the skewness are commonly represented by the Fano factor $%
F_{2}=C_{2}/C_{1}$ and $F_{3}=C_{3}/C_{1}$, respectively.

Since the electron-electron interaction may bring correlations and
entanglement of electron states, the shot noise in the mesoscopic
system with Coulomb interaction has attracted the significant attention\cite%
{Galaktionov,Golub}. The study of FCS of interacting electrons in
mesoscopic systems has become a challenging subject of great
interest \cite{Bagrets}. In this letter, we study the FCS of
electron transport through SMM weakly coupled to two metallic
electrodes, here the first three cumulants are given in terms of
numerical calculation. In the model considered here two electrodes
are regarded as non-interacting Fermi gases, while the central SMM
is treated as a multi-level system with finite Coulomb interaction.
We found that the effect of finite Coulomb interaction on the shot
noise depends not only on the internal level structure of the SMM,
but also on the left-right asymmetry of the SMM-electrode coupling.
In particular, our analytical results indicate that for finite
Coulomb interaction the FCS, which is different from the case of
$U\rightarrow \infty$, shows a symmetrical gate-voltage-dependence
when both the intensities of the SMM coupling to the left and right
electrodes are interchanged, which originates from the both
symmetries of the effective channel energy levels and the
probability distribution. Moreover, we also observed
super-Poissonian noise for symmetric coupling situation. The paper
is organized as follows. In Sec. II, we introduce the SMM system and
outline the procedure to obtain the FCS formalism based on a
particle-number-resolved quantum master equation. The numerical
results are discussed in Sec. III, where we analyze the
occurrence-mechanism of super-Poissonian noise and discuss the
effects of Coulomb interaction, the left-right asymmetry of
SMM-electrode coupling, and the applied gate voltage on the
super-Poissonian noise. Finally, in Sec. IV we summarize the work.

\section{MODEL AND FORMALISM}

A magnetic molecule coupled to two metallic electrodes L (left) and
R (right) is described by the Hamiltonian\cite{Timm} $%
H_{total}=H_{mol}+H_{Leads}+H_{T}.$ We assume that the SMM-electrode
coupling is sufficiently weak so that the electron transport is
dominated by
sequential tunneling through a single molecular level with on-site energy $%
\varepsilon_{d}$. The molecular Hamiltonian is given by
\begin{equation}
H_{mol}=(\varepsilon_{d}-eV_{g})\hat{n}+\frac{U}{2}\hat{n}(\hat{n}-1)-J\,%
\vec{s}\cdot\vec{S}-KS_{z}^{2}-B(s^{z}+S^{z}).  \label{model}
\end{equation}
Here the first two terms depict the lowest unoccupied molecular
orbital (LUMO), $\hat{n}\equiv
d_{\uparrow}^{\dag}d_{\uparrow}+d_{\downarrow}^{\dag
}d_{\downarrow}$ is the number operator, where $d_{\sigma}^{\dag}$ ($%
d_{\sigma}$) creates (annihilates) an electron with spin $\sigma$
and energy $\varepsilon_{d}$ (which can be tuned by applying a gate
voltage $V_{g}$) in the LUMO. $U$ is the Coulomb interaction between
two electrons in the LUMO. The third term describes the exchange
coupling between\ electron spin $\vec{s}$
in the LUMO and the giant spin $\vec{S}$, the electronic spin operator $\vec{%
s}\equiv \sum_{\sigma\sigma^{\prime}}d_{\sigma}^{\dag}\left( \vec{\sigma}%
_{\sigma\sigma^{\prime}}\right) d_{\sigma^{\prime}},$ where $\vec{\sigma }%
\equiv$ $(\sigma_{x},\sigma_{y},\sigma_{z})$ denotes the vector of
Pauli matrices. The forth term is the anisotropy energy of the
magnetic molecule whose easy-axis is $Z$-axis ($K>0$). The last term
denotes Zeeman splitting. For simplicity, we assume an external
magnetic field $B$ is applied along the easy axis of the SMM.

The relaxation in the electrodes is assumed to be sufficiently fast
so that their electron distributions can be described by equilibrium
Fermi functions. The electrodes are modeled as non-interacting Fermi
gases and the corresponding Hamiltonian
\begin{equation}
H_{Leads}=\sum_{\alpha \mathbf{k}\sigma }\varepsilon _{\alpha \mathbf{k}%
\sigma }a_{\alpha \mathbf{k}\sigma }^{\dag }a_{\alpha
\mathbf{k}\sigma }, \label{Leads}
\end{equation}%
where $a_{\alpha \mathbf{k}\sigma }^{\dag }$ ($a_{\alpha \mathbf{k}\sigma }$%
) creates (annihilates) an electron with energy $\varepsilon
_{\alpha
\mathbf{k}\sigma }$, momentum $\mathbf{k}$ and spin $\sigma $ in $\alpha $ ($%
\alpha =L,R$) electrode. The electron tunneling between the LUMO and
the electrodes is described by
\begin{equation}
H_{T}=\sum_{\alpha \mathbf{k}\sigma }\left( t_{\alpha }a_{\alpha \mathbf{k}%
\sigma }^{\dag }d_{\sigma }+H.c.\right) .  \label{tunneling}
\end{equation}%
According to Ref. \cite{Timm} the eigenstates of an isolated SMM
have four branches and may be denoted as $|n,m\rangle ^{\nu }$ where
$n$ ($n=0,1,2)$ is the electron number in the molecule orbital, and
$m$ is the quantum number for the $Z$-component of the total spin.
The index $\nu (=\pm )$ appears only when $n=1$. In term of the
electron state $\left\vert i,j\right\rangle _{LUMO}$
$(i,j=0,\uparrow ,\downarrow )$ in molecular
orbital and the local spin state $\left\vert m\right\rangle _{GS}$ ($m\in %
\left[ -S,S\right] $) the empty-branch and doubly-occupied states
may be expressed as
\begin{equation}
\left\vert 0,m\right\rangle =\left\vert 0,0\right\rangle
_{LUMO}\left\vert m\right\rangle _{GS},m\in \left[ -S,S\right] ,
\label{zero}
\end{equation}%
and%
\begin{equation}
\left\vert 2,m\right\rangle =\left\vert \uparrow ,\downarrow
\right\rangle _{LUMO}\left\vert m\right\rangle _{GS},m\in \left[
-S,S\right] ,  \label{two}
\end{equation}%
respectively, and the two singly-occupied branches are $\left\vert
1,\pm \left( S+1/2\right) \right\rangle =\left\vert \alpha _{\pm
}\right\rangle \left\vert \pm S\right\rangle $ ($\alpha
_{+}=\uparrow ,\alpha _{-}=\downarrow $) for $m=\pm \left(
S+1/2\right) $, and
\begin{equation}
\left\vert 1,m\right\rangle ^{\pm }=a_{m}^{\pm }\left\vert \uparrow
\right\rangle _{LUMO}\left\vert m-\frac{1}{2}\right\rangle
_{GS}+b_{m}^{\pm
}\left\vert \downarrow \right\rangle _{LUMO}\left\vert m+\frac{1}{2}%
\right\rangle _{GS},  \label{one}
\end{equation}%
with%
\begin{eqnarray*}
a_{m}^{\pm } &=&\frac{J\sqrt{S\left( S+1\right)
-m^{2}+1/4}}{2\sqrt{\Delta E\left( m\right) }\sqrt{2\Delta E\left(
m\right) \mp \left( 2K-J\right) m}},
\\
b_{m}^{\pm } &=&\mp \frac{\sqrt{2\Delta E\left( m\right) \mp \left(
2K-J\right) m}}{2\sqrt{\Delta E\left( m\right) }}.
\end{eqnarray*}%
for $-S+1/2\leq m\leq S-1/2,$ where $\Delta E\left( m\right) =\left[
K\left(
K-J\right) m^{2}+\left( J/4\right) ^{2}\left( 2S+1\right) ^{2}\right] ^{1/2}$%
. The corresponding energy eigenvalues of molecular eigenstates are given by%
\cite{Timm}%
\begin{equation}
\epsilon \left( 0,m\right) =-Km^{2}-Bm,  \label{energy0}
\end{equation}%
\begin{equation}
\epsilon \left( 2,m\right) =2\left( \varepsilon _{d}-eV_{g}\right)
+U-Km^{2}-Bm,  \label{energy2}
\end{equation}%
\begin{equation}
\epsilon ^{\pm }\left( 1,m\right) =\varepsilon _{d}-eV_{g}-Bm+\frac{J}{4}%
-K\left( m^{2}+\frac{1}{4}\right) \pm \Delta E\left( m\right) .
\label{energy1}
\end{equation}%
Here, for $\left\vert 1,\pm \left( S+1/2\right) \right\rangle $ the
upper (lower) sign applies if $K-J/2$ is positive (negative). It is
found from
Eqs. (\ref{energy0})--(\ref{energy1}) that the energy eigenvalues of $%
\left\vert 0,m\right\rangle $ is independent of gate voltage while those of $%
\left\vert 1,m\right\rangle ^{\pm }$ and $\left\vert
2,m\right\rangle $ depend on gate voltage, so that the internal
level structure can be tuned by an applied gate voltage. For the
finite Coulomb interaction case one need consider the contribution
of the doubly-occupied states $\left\vert 2,m\right\rangle $ to the
transport.

In sequential tunneling regime the transitions between the SMM and
the electrodes are well described by quantum master equation of a
reduced density matrix spanned by the eigenstates of the SMM. The
FCS of electron transport through the SMM may be implemented with
the help of the particle-number-resolved master equation for the
reduced density matrix which is given by\cite{Li1,Li2,WangSK}
\begin{equation}
\dot{\rho}^{\left( n\right) }\left( t\right)
=-i\mathcal{L}\rho^{\left( n\right) }\left( t\right)
-\frac{1}{2}\mathcal{R}\rho^{\left( n\right) }\left( t\right)
\label{Master1}
\end{equation}
under the second order Born approximation and Markovian
approximation, with Liouvillian superoperator $\mathcal{L}\left(
\cdots\right)=\left[ H_{mol},\left(\cdots\right)\right]$ and
\begin{eqnarray}
\mathcal{R}\rho^{\left(n\right)}\left(t\right)&=%
{\displaystyle\sum\limits_{\mu=\uparrow,\downarrow}}
\left[  d_{\mu}^{\dagger}A_{\mu}^{\left(  -\right)  }\rho^{\left(
n\right) }\left(  t\right)  +\rho^{\left(  n\right)  }\left(
t\right)  A_{\mu
}^{\left(  +\right)  }d_{\mu}^{\dagger}-A_{L\mu}^{\left(  -\right)  }%
\rho^{\left(  n\right)  }\left(  t\right)  d_{\mu}^{\dagger}\right.
\nonumber\\
&  \left.  -d_{\mu}^{\dagger}\rho^{\left(  n\right)  }\left(
t\right) A_{L\mu}^{\left(  +\right)  }-A_{R\mu}^{\left(  -\right)
}\rho^{\left(
n-1\right)  }\left(  t\right)  d_{\mu}^{\dagger}-d_{\mu}^{\dagger}%
\rho^{\left(  n+1\right)  }\left(  t\right)  A_{R\mu}^{\left(
+\right)
}\right]  +H.c..\label{Master2}%
\end{eqnarray}
Here $A_{\mu}^{\left( \pm\right)
}=\sum_{\alpha=L,R}A_{\alpha\mu}^{\left( \pm\right) }$,
$A_{\alpha\mu}^{\left( \pm\right)
}=\Gamma_{\alpha}n_{\alpha}^{\pm}\left( -\mathcal{L}\right) d_{\mu}$, $%
n_{\alpha}^{+}=f_{\alpha}$, $n_{\alpha}^{-}=1-f_{\alpha}$
($f_{\alpha}$ is the Fermi function of the electrode $\alpha$),
$\Gamma_{\alpha}=2\pi
g_{\alpha }\left\vert t_{\alpha}\right\vert ^{2}$ and $g_{\alpha}$ ($%
\alpha=R,L$) denotes the density of states of the metallic electrodes. $%
\rho^{\left( n\right) }\left( t\right) $ is the reduced density
matrix of the SMM conditioned by the electron numbers arriving at
the right electrode up to time $t$. Throughout this work, we set
$e\equiv\hbar =1$. The CGF connects with the
particle-number-resolved density matrix by defining $S\left(
\chi,t\right) =\sum_{n}\rho^{\left( n\right) }\left( t\right)
e^{in\chi}$. Evidently, we have $e^{-F\left( \chi\right)
}=$Tr$\left[ S\left( \chi,t\right) \right] $, where the trace is
evaluated over the eigenstates of the SMM. Since Eq. (\ref{Master1})
has the following form
\begin{equation}
\dot{\rho}^{\left( n\right) }=A\rho^{\left( n\right) }+C\rho^{\left(
n+1\right) }+D\rho^{\left( n-1\right) },  \label{formalmaster}
\end{equation}
then $S\left( \chi,t\right) $ satisfies
\begin{equation}
\mathcal{\dot{S}}=A\mathcal{S}+e^{-i\chi}C\mathcal{S}+e^{i\chi}D\mathcal{S}\equiv\mathcal{L}_{\chi}\mathcal{S},
\label{formalmaster1}
\end{equation}
where $\mathcal{S}$ is a column matrix, and A, C and D are three
square matrices. Here, the procedure for calculating the specific
form of $\mathcal{L}_{\chi}$ is given in detail in the Appendix. In
the low frequency limit, the counting time ($i.e.$, the time of
measurement) is much longer than the time of tunneling through the
SMM. In this case, $F\left( \chi\right) $ is given
by\cite{Bagrets,Groth}
\begin{equation}
F\left( \chi\right) =-\lambda_{1}\left( \chi\right) t,
\label{CGFformal}
\end{equation}
where $\lambda_{1}\left( \chi\right) $ is the eigenvalue of $\mathcal{L}%
_{\chi}$ which goes to zero for $\chi\rightarrow0$. According to the
definition of the cumulants one can express $\lambda_{1}\left( \chi\right) $%
\ as
\begin{equation}
\lambda_{1}\left( \chi\right) =\frac{1}{t}\sum_{k=1}^{\infty}C_{k}\frac{%
\left( i\chi\right) ^{k}}{k!}.  \label{Lambda}
\end{equation}
Inserting Eq. (\ref{Lambda}) into $\left\vert \mathcal{L}%
_{\chi}-\lambda_{1}\left( \chi\right) I\right\vert =0$ and expanding
this determinant in series of $\left( i\chi\right) ^{k}$, one can calculate $%
C_{k} $ by setting the coefficients of $i\chi$ equal to zero.

\section{NUMERICAL RESULTS AND DISCUSSION}

In the following we study the effect of finite Coulomb interaction
on the FCS of electronic transport through the SMM weakly coupled to
two metallic electrodes. We assume the bias voltage ($V_{b}=\mu
_{L}-\mu _{R}$) is symmetrically entirely dropped at the
SMM-electrode tunnel junctions, which implies that the levels of the
SMM are independent of the applied bias voltage even if the
couplings are not symmetric. The parameters of the SMM
are chosen as\cite{Timm} $S=2$, $\varepsilon _{d}=200\Gamma $, $J=100\Gamma $%
, $K=40\Gamma $, $B=80\Gamma $ and $k_{B}T=10\Gamma$, where $\Gamma
$\ is the typical rate for the tunneling of electrons between the
SMM and electrode. Here, the validity of the FCS formalism in this
work deserves some discussions. In the SMM system, the relaxation
rate of transition from a given molecular eigenstate $\left\vert n,m\right\rangle $%
of the SMM to the neighboring eigenstates $\left\vert n,m\pm
1\right\rangle $ of the same multiplet can be written as\cite%
{Misiorny}
\begin{equation}
\Gamma _{SMM}^{\left\vert n,m\right\rangle \rightarrow \left\vert
n,m\pm 1\right\rangle }=\frac{1}{\tau _{SMM}}\frac{1}{1+e^{-\left(
\epsilon _{\left\vert n,m\right\rangle }-\epsilon _{\left\vert
n,m\pm 1\right\rangle }\right) /k_{B}T}},  \label{Relaxation}
\end{equation}%
where $\tau _{SMM}$ denotes the SMM's spin-relaxation time, and
$\epsilon _{\left\vert n,m\right\rangle }$ the eigenvalue of the
molecular state $\left\vert n,m\right\rangle $. For a typical SMM at
low temperature (about the order of $1K$), e.g., Fe$_{8}$, $\tau
_{SMM}$ is of the order of $10^{-6} $ s\cite{TauSMM}, whereas the
tunneling time of electron through
the SMM $\tau _{0}$ of the order of $10^{-9}$ s\cite%
{Tauelec}. This means that an electron, which from
incident-electrode tunnels into the SMM, has enough time to tunnel
out the SMM before the corresponding molecular eigenstate relaxes to
its neighboring eigenstates of the same multiplet since $\tau
_{SMM}\gg $\textit{\ }$\tau _{0}$. Therefore, the effect of the
SMM's spin-relaxation processes on the FCS can be neglected. The
thermal energy corresponding to the temperature considered is much
smaller than energy barrier so that thermally activated transitions
above the barrier can be neglected. Furthermore, we assume the
external magnetic field is applied along the SMM's easy axis and
transverse anisotropy is small enough relative to the easy-axis
anisotropy so that the magnetic quantum tunneling can also be
neglected.

In the present work, we only study the transport above the
sequential tunneling threshold, i.e., $V_{b}>2\epsilon _{se}$, where
$\epsilon _{se}$\ is the energy difference between the ground state
with charge $N$\ and the first excited states $N-1$\cite{Aghassi1}.
In this regime, the inelastic sequential tunneling process is
dominant, thus electrons have sufficient energy to overcome the
Coulomb blockade and tunnel sequentially through the SMM. A special
emphasis is the effects of the strongly asymmetric coupling to two
electrodes and gate voltage on super-Poissonian noise. The
sequential tunneling requires a change of the electron number by
$\Delta n=\pm 1$ and the magnetic quantum number by $\Delta m=\pm
1/2$, so that the sequential tunneling threshold depends on the
effective channel energy levels in bias voltage window which are
given by

\begin{equation}
\epsilon_{1}^{\pm}\left( \uparrow\right) =\epsilon^{\pm}(1,m+\frac{1}{2}%
)-\epsilon(0,m)=\epsilon_{d}-V_{g}+E^{\pm}(m+\frac{1}{2}),
\label{level1}
\end{equation}

\begin{equation}
\epsilon _{2}^{\pm }\left( \downarrow \right) =\epsilon
(2,m)-\epsilon ^{\pm }(1,m+\frac{1}{2})=\epsilon _{d}-V_{g}+U-E^{\pm
}(m+\frac{1}{2}), \label{level2}
\end{equation}%
\begin{equation}
\epsilon _{3}^{\pm }\left( \uparrow \right) =\epsilon (2,m)-\epsilon
^{\pm }(1,m-\frac{1}{2})=\epsilon _{d}-V_{g}+U-E^{\pm
}(m-\frac{1}{2}), \label{level3}
\end{equation}%
\begin{equation}
\epsilon _{4}^{\pm }\left( \downarrow \right) =\epsilon ^{\pm }(1,m-\frac{1}{%
2})-\epsilon (0,m)=\epsilon _{d}-V_{g}+E^{\pm }(m-\frac{1}{2}),
\label{level4}
\end{equation}%
where $E^{\pm }(m+1/2)=-\left( m+1/2\right) K-B/2+J/4\pm \Delta
E(m+1/2)$ and $E^{\pm }(m-1/2)=\left( m-1/2\right) K+B/2+J/4\pm
\Delta E(m-1/2)$. The electron with spin $\sigma (=\uparrow
,\downarrow )$ can be transferred by the effective channel energy levels $%
\epsilon _{i}^{\pm }(\sigma )(i=1,2,3,4)$. The presence of finite
Coulomb interaction may induce that the effective channel energy
levels move symmetrically in the bias voltage window with increasing
the bias voltage. From Eqs. (\ref{level1})--(\ref{level4}), one may
find $\left\vert \epsilon _{1}^{\pm }(\uparrow )\right\vert
=\left\vert \epsilon _{2}^{\pm }\left( \downarrow \right)
\right\vert $ and $\left\vert \epsilon _{3}^{\pm }\left( \uparrow
\right) \right\vert =\left\vert \epsilon _{4}^{\pm }\left(
\downarrow \right) \right\vert $ when $V_{g}=\epsilon _{d}+U/2,$
which indicates the effective channel energy levels $\epsilon
_{1}^{\pm }\left(
\downarrow \right) $ and $\epsilon _{2}^{\pm }\left( \uparrow \right) $, $%
\epsilon _{3}^{\pm }\left( \downarrow \right) $ and $\epsilon
_{4}^{\pm }\left( \uparrow \right) $ enter synchronously the bias
voltage window with
increasing the bias voltage. When the gate voltage has a departure from $%
V_{g}=\epsilon _{d}+U/2$, for example, $V_{g}=\epsilon _{d}+U/2\pm
\Delta
V $, the effective channel energy levels still have similar symmetry,%
\begin{equation}
\left\vert \epsilon _{1}^{\pm }(\uparrow )\right\vert
(V_{g}=\epsilon _{d}+U/2\pm \Delta V)=\left\vert \epsilon _{2}^{\pm
}\left( \downarrow \right) \right\vert (V_{g}=\epsilon _{d}+U/2\mp
\Delta V), \label{symmetry1}
\end{equation}%
\begin{equation}
\left\vert \epsilon _{3}^{\pm }(\downarrow )\right\vert
(V_{g}=\epsilon _{d}+U/2\pm \Delta V)=\left\vert \epsilon _{4}^{\pm
}\left( \uparrow \right) \right\vert (V_{g}=\epsilon _{d}+U/2\mp
\Delta V).  \label{symmetry2}
\end{equation}%
For the case of $U=100\Gamma $, the current as a function of the bias voltage for $%
\Gamma _{L}/\Gamma _{R}=10$ and $0.1$ is shown in Figs. 1(a) and
(d). Both
currents for the gate voltage $V_{g}=100\Gamma $ and $400\Gamma $ $%
(V_{g}=200\Gamma $ and $300\Gamma )$ have same sequential tunneling
threshold, which corresponds to $\Delta V=150\Gamma $ ($\Delta
V=50\Gamma $). In particular, we find that the finite Coulomb
interaction may induce that the FCS shows the same
bias-voltage-dependence under different external conditions. Figures
1(a) and (d), (b) and (e), and (c) and (f) show the average current,
shot noise and skewness for the two different strong asymmetric
couplings ($\Gamma _{L}/\Gamma _{R}=10$ and $0.1$) in the presence
of finite Coulomb interaction ($U=100\Gamma$), respectively. It is
interesting to note that the first three cumulants for $%
\Gamma _{L}/\Gamma _{R}=10$ at $V_{g}=100\Gamma $, $200\Gamma $, $%
250\Gamma $, $300\Gamma $ and $400\Gamma $ have the same
bias-voltage-dependence as that for $\Gamma _{L}/\Gamma _{R}=0.1$ at
$V_{g}=400\Gamma $, $300\Gamma $, $250\Gamma $, $200\Gamma $ and
$100\Gamma $, respectively. In fact,
besides the symmetry of the effective channel energy levels in Eqs. (\ref%
{symmetry1}) and (\ref{symmetry2}), Fermi distribution functions
also satisfy the relation

\begin{equation}
\left. f_{L/R}[\epsilon_{2}^{\pm}(\downarrow)]\right\vert
_{V_{g}=\epsilon_{d}+U/2\pm\Delta V}=1-\left.
f_{R/L}[\epsilon_{1}^{\pm}(\uparrow)]\right\vert
_{V_{g}=\epsilon_{d}+U/2\mp\Delta V},  \label{condition1}
\end{equation}

\begin{equation}
\left. f_{L/R}[\epsilon_{3}^{\pm}(\downarrow)]\right\vert
_{V_{g}=\epsilon_{d}+U/2\pm\Delta V}=1-\left.
f_{R/L}[\epsilon_{4}^{\pm}(\uparrow)]\right\vert
_{V_{g}=\epsilon_{d}+U/2\mp\Delta V},  \label{condition2}
\end{equation}
which may ensure

\begin{equation}
\left. \left\langle m,2\right\vert \dot{\rho}\left\vert
2,m\right\rangle \right\vert _{\Gamma _{L}=\Gamma _{1},\Gamma
_{R}=\Gamma _{2}}=\left. \left\langle m,0\right\vert
\dot{\rho}\left\vert 0,m\right\rangle \right\vert _{\Gamma
_{L}=\Gamma _{2},\Gamma _{R}=\Gamma _{1}},
\end{equation}%
\begin{equation}
\left. ^{\pm }\left\langle m,1\right\vert \dot{\rho}\left\vert
1,m\right\rangle ^{\pm }\right\vert _{\Gamma _{L}=\Gamma _{1},\Gamma
_{R}=\Gamma _{2}}=\left. ^{\pm }\left\langle m,1\right\vert \dot{\rho}%
\left\vert 1,m\right\rangle ^{\pm }\right\vert _{\Gamma _{L}=\Gamma
_{2},\Gamma _{R}=\Gamma _{1}},
\end{equation}%
where $\rho \left( t\right) =\sum_{n}\rho ^{\left( n\right) }\left(
t\right) $. Based on these conditions one may prove the probability
distribution
\begin{equation}
\left. P_{\left\vert 2,m\right\rangle }\right\vert _{\Gamma
_{L}=\Gamma _{1},\Gamma _{R}=\Gamma _{2}}=\left. P_{\left\vert
0,m\right\rangle }\right\vert _{\Gamma _{L}=\Gamma _{2},\Gamma
_{R}=\Gamma _{1}}, \label{probabilty1}
\end{equation}%
\begin{equation}
\left. P_{\left\vert 1,m\right\rangle ^{\pm }}\right\vert _{\Gamma
_{L}=\Gamma _{1},\Gamma _{R}=\Gamma _{2}}=\left. P_{\left\vert
1,m\right\rangle ^{\pm }}\right\vert _{\Gamma _{L}=\Gamma
_{2},\Gamma _{R}=\Gamma _{1}},  \label{probabilty2}
\end{equation}%
whose numerical results are shown in Fig. 2. It is because of the
both symmetries of the effective channel energy levels and the
probability distributions that the FCS of transport through the SMM
for $\Gamma _{L}/\Gamma _{R}=10$ at $V_{g}=\epsilon
_{d}+U/2\pm \Delta V$ represents the same bias-voltage-dependence as that for $%
\Gamma _{L}/\Gamma _{R}=0.1$ at $V_{g}=\epsilon _{d}+U/2\mp \Delta
V$,
which are shown in Fig. 1 for a given finite Coulomb interaction $%
U=100\Gamma $. The transport characteristic should easily be
observed experimentally by reversing the bias voltage between the
left and right
electrodes and tuning the applied gate voltage; but for the case of $%
U\rightarrow \infty $, this feature will not be observed because the
transition processes between doubly-occupied and the singly-occupied
states are prohibited.

It is generally thought that the Coulomb interaction gives rise to
the suppression of shot noise\cite{Reznikov,Hamasaki}. For the
present SMM system with finite Coulomb interaction, however, the
super-Poissonian shot noise is observed in the sequential tunneling
regime. For a given $U=100\Gamma $, as shown in Figs. 1(b) and (e),
the super-Poissonian noise for $\Gamma _{L}/\Gamma _{R}=10$ occurs
at higher gate voltage such as $V_{g}=300\Gamma $ and $400\Gamma $,
while that for $\Gamma _{L}/\Gamma _{R}=0.1$ occurs at lower gate
voltage such as $V_{g}=200\Gamma $ and $100\Gamma $, which is
consistent with the $U\rightarrow \infty $ case\cite{Xue}. The
presence of the super-Poissonian noise in the sequential tunneling
regime can be understood with the help of the dynamic competition
between effective fast and slow
channels\cite{Xue,Aguado,WangSK,Aghassi1,Safonov,Djuric,Aghassi2}.
In order to give a qualitative explanation, we plot the six main
molecular channel currents for $V_{g}=400\Gamma $ as a function of
bias voltage $V_{b}$ in Fig. 3(a). Here, the calculation of the
molecular channel currents can be found in Refs. \cite{Timm,Xue}.
When the bias voltage increases up to about $170\Gamma $, the fast
channel current $I_{\left\vert 2,2\right\rangle \longrightarrow
\left\vert 1,5/2\right\rangle }$ begins to compete with the slow
channel currents $I_{\left\vert 2,-2\right\rangle \rightarrow
\left\vert 1,-5/2\right\rangle }$, $I_{\left\vert 2,\pm
2\right\rangle \rightarrow \left\vert 1,\pm 3/2\right\rangle ^{-}}$ and $%
I_{\left\vert 2,\pm 1\right\rangle \rightarrow \left\vert 1,\pm
3/2\right\rangle ^{-}}$, but the competition is quickly destroyed
due to the sum of slow channel currents being approach to the fast
channel current, the corresponding noise [the short dashed line in
Fig. 1(b)] is only enhanced but does not reach the super-Poissonian.
With further increase of the bias voltage (about $250\Gamma $),
three sets of the fast-and-slow channel currents are formed, $i.e.$,
$I_{\left\vert 2,\pm 2\right\rangle \rightarrow \left\vert 1,\pm
5/2\right\rangle}$ , $I_{\left\vert 2,\pm
2\right\rangle \rightarrow \left\vert 1,\pm 3/2\right\rangle ^{-}}$ and $%
I_{\left\vert 2,\pm 1\right\rangle \rightarrow \left\vert 1,\pm
3/2\right\rangle ^{-}}$ , which result in the second enhancement of
shot noise. When the bias increases to about $500\Gamma $, the fast
channel currents ($I_{\left\vert 2,2\right\rangle \longrightarrow
\left\vert 1,5/2\right\rangle }$, $I_{\left\vert 2,2\right\rangle
\rightarrow \left\vert 1,3/2\right\rangle ^{-}}$ and $I_{\left\vert
2,1\right\rangle \rightarrow \left\vert 1,3/2\right\rangle ^{-}}$)
begin to decrease, while the corresponding slow channel currents
($I_{\left\vert 2,-2\right\rangle \longrightarrow \left\vert
1,-5/2\right\rangle }$, $I_{\left\vert
2,-2\right\rangle \rightarrow \left\vert 1,-3/2\right\rangle ^{-}}$ and $%
I_{\left\vert 2,-1\right\rangle \rightarrow \left\vert
1,-3/2\right\rangle ^{-}}$) begin to increase. The competition
between the increase of the slow channel currents and the decrease
of the fast channel currents finally leads
to appearing of super-Poissonian Fano factor in the bias from about $%
500\Gamma $ to $625\Gamma $ [the short dashed line in Fig. 1(b)].
Furthermore, we also observe that the super-Poissonian distribution
of the skewness $F_{3}$ in the sequential tunneling regime [Fig.
1(c)], which seems sensitive only to the competition between the
fast channels of current
decreasing and the slow channels of current increasing [Fig. 3(a)]. As for $%
\Gamma _{L}/\Gamma _{R}=0.1$\ at lower gate voltage, the mechanism
of shot-noise enhancement originates from the same reason but the
effective fast-and-slow transport channels consist of the
transitions from the singly-occupied states to empty states, see
Fig. 3(b). Furthermore, in the Coulomb blockade regime (see Fig. 1),
the shot noise enhancement, as explained in Ref. \cite{Aghassi1}, is
due to the possible thermal occupation and subsequent sequential
depletion of excited states that lead to small cascades of tunneling
events interrupted by long Coulomb blockages. In fact, since in the
Coulomb blockade regime the current is exponentially suppressed and
the electron transport is dominated by cotunneling, when taking into
account cotunneling the normalized second and third moments will
deviate from the results obtained by only sequential
tunneling\cite{Thielmann}.

In contrast with the above case, we find that the finite Coulomb
interaction plays a crucial role in determining whether the
super-Poissonian noise occurs in
situations with relatively small gate voltage for $\Gamma _{L}/\Gamma _{R}>1$%
\ and relatively large gate voltage for $\Gamma _{L}/\Gamma _{R}<1$.
In order to study the effects of finite Coulomb interaction on the
shot noise in transport through the SMM for the two above-mentioned
cases, Fig. 4 shows the current and the shot noise as a function of
the bias voltage for various Coulomb interaction energies. For the
case of $U\rightarrow \infty $, the super-Poissonian noise, as shown
in Figs. 4(b) and (d), dose not appear in the sequential tunneling
regime (also see Fig. 2 in Ref. \cite{Xue}). Here, for the case of
$\Gamma _{L}/\Gamma _{R}=0.1$, $U\rightarrow \infty $\ and
$V_{g}=300\Gamma $, it should be noted that the super-Poissonian
noise occurs in the Coulomb blockade regime, see the solid line in
Fig. 4(d). Comparing with the $U\rightarrow \infty $\ case, the
finite Coulomb interaction may induce the super-Poissonian noise
only when it is larger than a certain value. We take the case of
$V_{g}=100\Gamma $\ and $\Gamma _{L}/\Gamma _{R}=10$\ for
illustration. For small Coulomb interaction (for example,
$U=100\Gamma $) the super-Poissonian shot noise does not appear
while for relatively larger Coulomb interaction (for example,
$U\gtrsim 200\Gamma $) the super-Poissonian shot noise is observed
[see Fig. 4(b)]. The role of Coulomb interaction in enhancing shot
noise can be understood with the help of the main channel currents
shown in Fig. 5. The currents in Figs. 5(a) and (c) result from the
electron transitions from singly occupied to empty states while
Figs. (b) and (d) correspond to the transitions from doubly occupied
to singly occupied states. For small Coulomb interaction
$U=100\Gamma $, when the bias changes from $200\Gamma $ to
$500\Gamma $ the competition between the fast channel current
($I_{\left\vert 1,5/2\right\rangle \longrightarrow \left\vert
0,2\right\rangle })$ and the slow channel currents ($I_{\left\vert
1,-5/2\right\rangle \longrightarrow \left\vert 0,-2\right\rangle }$,
$I_{\left\vert 1,\pm 3/2\right\rangle ^{-}\longrightarrow \left\vert
0,\pm 2\right\rangle }$ and $I_{\left\vert
1,\pm 3/2\right\rangle ^{-}\longrightarrow \left\vert 0,\pm 1\right\rangle }$%
) is an active competition in which the increase (or decrease) of
the fast channel current is always accompanied by the decrease (or
increase) of the slow channel currents, which makes the shot noise
increase from 0.84 to 0.93 [see Fig. 4(b)].
With the bias approaching to $600\Gamma $ the currents $%
I_{\left\vert 2,2\right\rangle \longrightarrow \left\vert
1,5/2\right\rangle }$ and $I_{\left\vert 2,2\right\rangle
\longrightarrow \left\vert 1,3/2\right\rangle ^{\pm }}$ begin to
compete with $I_{\left\vert 1,5/2\right\rangle \longrightarrow
\left\vert 0,2\right\rangle }$, but the rapid increase of the
amplitudes of $I_{\left\vert 2,2\right\rangle \longrightarrow
\left\vert 1,5/2\right\rangle }$ and $I_{\left\vert 2,2\right\rangle
\longrightarrow \left\vert 1,3/2\right\rangle ^{\pm }}$ lead to
rapid decrease of $I_{\left\vert 1,5/2\right\rangle \longrightarrow
\left\vert 0,2\right\rangle }$ and even destroy the competition, as
a result the shot noise quickly drops to sub-Poissonian after
increasing to 0.95 near
$V_{b}=640\Gamma $[see fig. 4(b)]. When Coulomb interaction increases to $%
200\Gamma $, the bias voltage regime in which the active competition
occurs is much larger than that for $U=100\Gamma$, $i.e.$, from $200\Gamma $ to $%
750\Gamma $ [see Fig. 5(c)], so that the shot noise has reached
super-Poissonian before the effective competition between the fast
and slow channels is destroyed by the rapid rise of $I_{\left\vert
2,2\right\rangle \longrightarrow \left\vert 1,5/2\right\rangle }$
and $I_{\left\vert 2,2\right\rangle \longrightarrow \left\vert
1,3/2\right\rangle ^{\pm }}$ [see Figs. 5(c) and (d)].

On the other hand, although the shot noise of finite Coulomb
interaction for $\Gamma _{L}/\Gamma _{R}=0.1$ in Fig. 4(d) have the
same bias-voltage-dependence as that for $\Gamma _{L}/\Gamma
_{R}=10$ in Fig. 4(b), the corresponding transport processes are
different due to occurring at different gate voltages, which may be
found by comparing
the main channel currents in Fig. 6(a) (for $\Gamma _{L}/\Gamma _{R}=0.1$, $%
U=100\Gamma $ and $V_{g}=400\Gamma $) with those in Figs. 5(a) and (b) (for $%
\Gamma _{L}/\Gamma _{R}=10,$ $U=100\Gamma $ and $V_{g}=100\Gamma $).
In particular, for the case of $\Gamma _{L}/\Gamma _{R}=0.1,$
$U=100\Gamma $ and $V_{g}=400\Gamma $ there is a reverse current
$I_{\left\vert
0,2\right\rangle \longrightarrow \left\vert 1,5/2\right\rangle }$ near $%
V_{b}=600\Gamma $ besides a rapidly increasing current
$I_{\left\vert 1,5/2\right\rangle \longrightarrow \left\vert
0,2\right\rangle }$ [see Fig. 6(a)] and the two currents destroy the
competition between the fast and slow
channels, resulting in rapid decrease of the shot noise. For the case of $%
U=200\Gamma $ and $V_{g}=500\Gamma $, as shown in Fig. 6(b), the
bias voltage regime in which the active competition occurs is much
larger than that for $U=100\Gamma $ and $V_{g}=400\Gamma ,$ $i.e.,$
from $200\Gamma $ to $750\Gamma $ (it is from $200\Gamma $ to
$640\Gamma $ for $U=100\Gamma $ and $V_{g}=400\Gamma $), so that the
shot noise has reached super-Poissonian before the reverse current
and the rapidly increasing current occur. As a result the shot noise
rapidly drops from the super-Poissonian to the sub-Poissonian.

The theoretical investigations have predicted that the
super-Poissonian noise may occur in the quantum dot
systems\cite{Djuric,Aghassi2,Thielmann} and molecular
junction\cite{Welack} coupled symmetrically to two electrodes. For
the present SMM system, when it symmetrically couples to two leads
the super-Poissonian noise in the sequential tunneling regime is
observed only under an appropriate finite Coulomb interaction that
determined by the parameters of the SMM (see Fig. 7), which still
not appears in the case of $U=0$\ and $U\rightarrow \infty $. The
noise characteristics can still be attributed to the competition
between the effective fast and slow transport channels.

\section{Conclusions}

We have studied the effect of finite Coulomb interaction on the FCS
of electron transport through a SMM weakly coupled to two metallic
electrodes above the sequential tunneling threshold by means of the
particle-number-resolved quantum master equation. The electron
transport through the SMM with finite Coulomb interaction $U$ has
richer current correlation due to the transitions of doubly-occupied
to singly-occupied states participating in transport. Compared with
the $U\rightarrow \infty $ case, our analytical results show that
the finite Coulomb interaction induces that the effective channel
energy levels move symmetrically into the bias voltage window with
increasing the bias voltage, as a result the FCS for $%
V_{g}=\epsilon _{d}+U/2\pm \Delta V$ has the same
bias-voltage-dependence as that for $V_{g}=\epsilon _{d}+U/2\mp
\Delta V$ (i.e., the FCS shows the symmetrical
gate-voltage-dependence) when both the intensities of the SMM
coupling to the left and right electrodes are interchanged, which
should easily be observed experimentally by reversing the bias
voltage. Moreover, we find that the effect of finite Coulomb
interaction on the shot noise depends not only on the gate voltage,
but also on the left-right asymmetry of SMM-electrode couplings. In
the case of $\Gamma _{L}/\Gamma_{R}>1$, for an arbitrary
given $U$\ (which contains $U=0$\ and $%
U\rightarrow \infty $) the super-Poissonian shot noise in the
sequential tunneling regime may be observed at a relatively large
gate voltage; whereas at a relatively small gate voltage the
super-Poissonian shot noise does not occur for the cases of $U=0$\
and $U\rightarrow \infty $, and may appear only when $U$ is the
finite value which is related to the parameters of the SMM. For the
$\Gamma _{L}/\Gamma _{R}<1$ case, the super-Poissonian shot noise is
found at a relatively small gate voltage for an arbitrary given $U$;
whereas at a relatively large gate voltage it may occur only for a
relatively large finite Coulomb interaction, which is contrary to
the $\Gamma _{L}/\Gamma _{R}>1 $ case. These characteristics of shot
noise can be understood as a result of the active competition
between the fast and slow channel currents.

\section*{Acknowledgments}

This work was supported by the Graduate Outstanding Innovation Item
of Shanxi Province (Grant No. 20103001), the National Nature Science
Foundation of China (Grant No. 10774094, No. 10775091 and No.
10974124) and the Shanxi Nature Science Foundation of China (Grant
No. 2009011001-1 and No. 2008011001-2).

\section*{Appendix}
\renewcommand{\theequation}{A.\arabic{equation}}
\setcounter{equation}{0}

In this appendix, we take $\left\langle 1,-\frac{5}{2}\right\vert
S\left( \chi,t\right) \left\vert 1,-\frac{5}{2}\right\rangle $ as an
example to
illustrate the procedure for obtaining the specific form of $\mathcal{L}%
_{\chi}$. Based on the $S\left( \chi,t\right)$ definition and Eq. (\ref%
{Master1}), after a careful calculation, the equation of motion for $%
\left\langle 1,-\frac{5}{2}\right\vert S\left( \chi,t\right)\left\vert 1,-%
\frac{5}{2}\right\rangle $ is given by

\begin{eqnarray}
&&\left\langle 1,-\frac{5}{2}\right\vert \overset{\cdot }{S}\left(
\chi
,t\right) \left\vert 1,-\frac{5}{2}\right\rangle   \notag \\
&=&-\Gamma _{L}n_{L}^{\left( +\right) }\left( \epsilon _{\left\vert
2,-2\right\rangle }-\epsilon _{\left\vert
1,-\frac{5}{2}\right\rangle
}\right) \left\langle 1,-\frac{5}{2}\right\vert S\left\vert 1,-\frac{5}{2}%
\right\rangle   \notag \\
&&-\Gamma _{R}n_{R}^{\left( +\right) }\left( \epsilon _{\left\vert
2,-2\right\rangle }-\epsilon _{\left\vert
1,-\frac{5}{2}\right\rangle
}\right) \left\langle 1,-\frac{5}{2}\right\vert S\left\vert 1,-\frac{5}{2}%
\right\rangle   \notag \\
&&-\Gamma _{L}n_{L}^{\left( -\right) }\left( \epsilon _{\left\vert 1,-\frac{5%
}{2}\right\rangle }-\epsilon _{\left\vert 0,-2\right\rangle }\right)
\left\langle 1,-\frac{5}{2}\right\vert S\left\vert 1,-\frac{5}{2}%
\right\rangle   \notag \\
&&-\Gamma _{R}n_{R}^{\left( -\right) }\left( \epsilon _{\left\vert 1,-\frac{5%
}{2}\right\rangle }-\epsilon _{\left\vert 0,-2\right\rangle }\right)
\left\langle 1,-\frac{5}{2}\right\vert S\left\vert 1,-\frac{5}{2}%
\right\rangle   \notag \\
&&+\Gamma _{L}n_{L}^{\left( -\right) }\left( \epsilon _{\left\vert
2,-2\right\rangle }-\epsilon _{\left\vert
1,-\frac{5}{2}\right\rangle }\right) \left\langle 2,-2\right\vert
S\left\vert 2,-2\right\rangle   \notag
\\
&&+\Gamma _{R}n_{R}^{\left( -\right) }\left( \epsilon _{\left\vert
2,-2\right\rangle }-\epsilon _{\left\vert
1,-\frac{5}{2}\right\rangle }\right) e^{i\chi }\left\langle
2,-2\right\vert S\left\vert
2,-2\right\rangle   \notag \\
&&+\Gamma _{L}n_{L}^{\left( +\right) }\left( \epsilon _{\left\vert 1,-\frac{5%
}{2}\right\rangle }-\epsilon _{\left\vert 0,-2\right\rangle }\right)
\left\langle 0,-2\right\vert S\left\vert 0,-2\right\rangle   \notag \\
&&+\Gamma _{R}n_{R}^{\left( +\right) }\left( \epsilon _{\left\vert 1,-\frac{5%
}{2}\right\rangle }-\epsilon _{\left\vert 0,-2\right\rangle }\right)
e^{-i\chi }\left\langle 0,-2\right\vert S\left\vert
0,-2\right\rangle . \label{A.1}
\end{eqnarray}%
Similarly, we can carry out the equations of motion for other matrix
elements, which can be rewritten as a compact matrix form%
\begin{equation}
\mathcal{\dot{S}}=\mathcal{L}_{\chi }\mathcal{S}.  \label{A.2}
\end{equation}%
Here, the column matrix $\mathcal{S}$ has the following form
\begin{eqnarray}
\mathcal{S}&=&\left[ \left\langle 1,-\frac{5}{2}\right\vert
S\left\vert 1,-\frac{5}{2}\right\rangle ,\left\langle
2,-2\right\vert
S\left\vert 2,-2\right\rangle ,\right.   \notag \\
&&\left. \left\langle 0,-2\right\vert S\left\vert 0,-2\right\rangle
,^{+}\left\langle 1,-\frac{3}{2}\right\vert S\left\vert 1,-\frac{3}{2}%
\right\rangle ^{+},^{+}\left\langle 1,-\frac{3}{2}\right\vert S\left\vert 1,-%
\frac{3}{2}\right\rangle ^{-},\right.   \notag \\
&&\left. ^{-}\left\langle 1,-\frac{3}{2}\right\vert S\left\vert 1,-\frac{3}{2%
}\right\rangle ^{+},^{-}\left\langle 1,-\frac{3}{2}\right\vert
S\left\vert 1,-\frac{3}{2}\right\rangle ^{-},\left\langle
2,-1\right\vert S\left\vert
2,-1\right\rangle ,\right.   \notag \\
&&\left. \left\langle 0,-1\right\vert S\left\vert 0,-1\right\rangle
,^{+}\left\langle 1,-\frac{1}{2}\right\vert S\left\vert 1,-\frac{1}{2}%
\right\rangle ^{+},^{+}\left\langle 1,-\frac{1}{2}\right\vert S\left\vert 1,-%
\frac{1}{2}\right\rangle ^{-},\right.   \notag \\
&&\left. ^{-}\left\langle 1,-\frac{1}{2}\right\vert S\left\vert 1,-\frac{1}{2%
}\right\rangle ^{+},^{-}\left\langle 1,-\frac{1}{2}\right\vert
S\left\vert 1,-\frac{1}{2}\right\rangle ^{-},\left\langle
2,0\right\vert S\left\vert
2,0\right\rangle ,\right.   \notag \\
&&\left. \left\langle 0,0\right\vert S\left\vert 0,0\right\rangle
,^{+}\left\langle 1,\frac{1}{2}\right\vert S\left\vert 1,\frac{1}{2}%
\right\rangle ^{+},^{+}\left\langle 1,\frac{1}{2}\right\vert S\left\vert 1,%
\frac{1}{2}\right\rangle ^{-},\right.   \notag \\
&&\left. ^{-}\left\langle 1,\frac{1}{2}\right\vert S\left\vert 1,\frac{1}{2}%
\right\rangle ^{+},^{-}\left\langle 1,\frac{1}{2}\right\vert S\left\vert 1,%
\frac{1}{2}\right\rangle ^{-},\left\langle 2,1\right\vert
S\left\vert
2,1\right\rangle ,\right.   \notag \\
&&\left. \left\langle 0,1\right\vert S\left\vert 0,1\right\rangle
,^{+}\left\langle 1,\frac{3}{2}\right\vert S\left\vert 1,\frac{3}{2}%
\right\rangle ^{+},^{+}\left\langle 1,\frac{3}{2}\right\vert S\left\vert 1,%
\frac{3}{2}\right\rangle ^{-},\right.   \notag \\
&&\left. ^{-}\left\langle 1,\frac{3}{2}\right\vert S\left\vert 1,\frac{3}{2}%
\right\rangle ^{+},^{-}\left\langle 1,\frac{3}{2}\right\vert S\left\vert 1,%
\frac{3}{2}\right\rangle ^{-},\left\langle 2,2\right\vert
S\left\vert
2,2\right\rangle ,\right.   \notag \\
&&\left. \left\langle 0,2\right\vert S\left\vert 0,2\right\rangle
,\left\langle 1,\frac{5}{2}\right\vert S\left\vert 1,\frac{5}{2}%
\right\rangle \right] ^{T}.  \label{A.3}
\end{eqnarray}%
From Eq. (\ref{A.1}), the first row of $\mathcal{L}_{\chi }$ is given by%
\begin{eqnarray}
\left[ {\mathcal{L}_{\chi }}\right] _{1\times 28} &=&\left[ \left[
{\mathcal{L}_{\chi }}\right]_{1,1},\right.
\notag \\
&&\left. \left[ {\mathcal{L}_{\chi }}\right]_{1,2},\left[
{\mathcal{L}_{\chi }}\right]_{1,3},0,0,0,0,0,0,\right.   \notag \\
&&\left. 0,0,0,0,0,0,0,0,\right.   \notag \\
&&\left. 0,0,0,0,0,0,0,0,\right.   \notag \\
&&\left. 0,0,0\right]   \label{A.4}
\end{eqnarray}%
with%
\begin{eqnarray*}
\left[ {\mathcal{L}_{\chi }}\right]_{1,1} &=&-\Gamma
_{L}n_{L}^{\left( +\right) }\left( \epsilon _{\left\vert
2,-2\right\rangle }-\epsilon _{\left\vert
1,-\frac{5}{2}\right\rangle }\right) -\Gamma _{R}n_{R}^{\left(
+\right) }\left( \epsilon _{\left\vert 2,-2\right\rangle
}-\epsilon _{\left\vert 1,-\frac{5}{2}\right\rangle }\right)  \\
&&-\Gamma _{L}n_{L}^{\left( -\right) }\left( \epsilon _{\left\vert 1,-\frac{5}{2}%
\right\rangle }-\epsilon _{\left\vert 0,-2\right\rangle }\right)
-\Gamma _{R}n_{R}^{\left( -\right) }\left( \epsilon _{\left\vert
1,-\frac{5}{2}\right\rangle }-\epsilon _{\left\vert
0,-2\right\rangle }\right) ,
\end{eqnarray*}%
\begin{equation*}
\left[ {\mathcal{L}_{\chi }}\right]_{1,2}=\Gamma _{L}n_{L}^{\left(
-\right) }\left( \epsilon _{\left\vert 2,-2\right\rangle }-\epsilon
_{\left\vert 1,-\frac{5}{2}\right\rangle }\right) +\Gamma
_{R}n_{R}^{\left( -\right) }\left( \epsilon _{\left\vert
2,-2\right\rangle }-\epsilon _{\left\vert
1,-\frac{5}{2}\right\rangle }\right) e^{i\chi },
\end{equation*}%
\begin{equation*}
\left[ {\mathcal{L}_{\chi }}\right]_{1,3}=\Gamma _{L}n_{L}^{\left( +\right) }\left( \epsilon _{\left\vert 1,-\frac{5}{2}%
\right\rangle }-\epsilon _{\left\vert 0,-2\right\rangle }\right)
+\Gamma _{R}n_{R}^{\left( +\right) }\left( \epsilon _{\left\vert
1,-\frac{5}{2}\right\rangle }-\epsilon _{\left\vert
0,-2\right\rangle }\right) e^{-i\chi }.
\end{equation*}

\newpage

\newpage

\begin{figure}[t]
\begin{center}
\includegraphics[height=10cm,width=12cm]{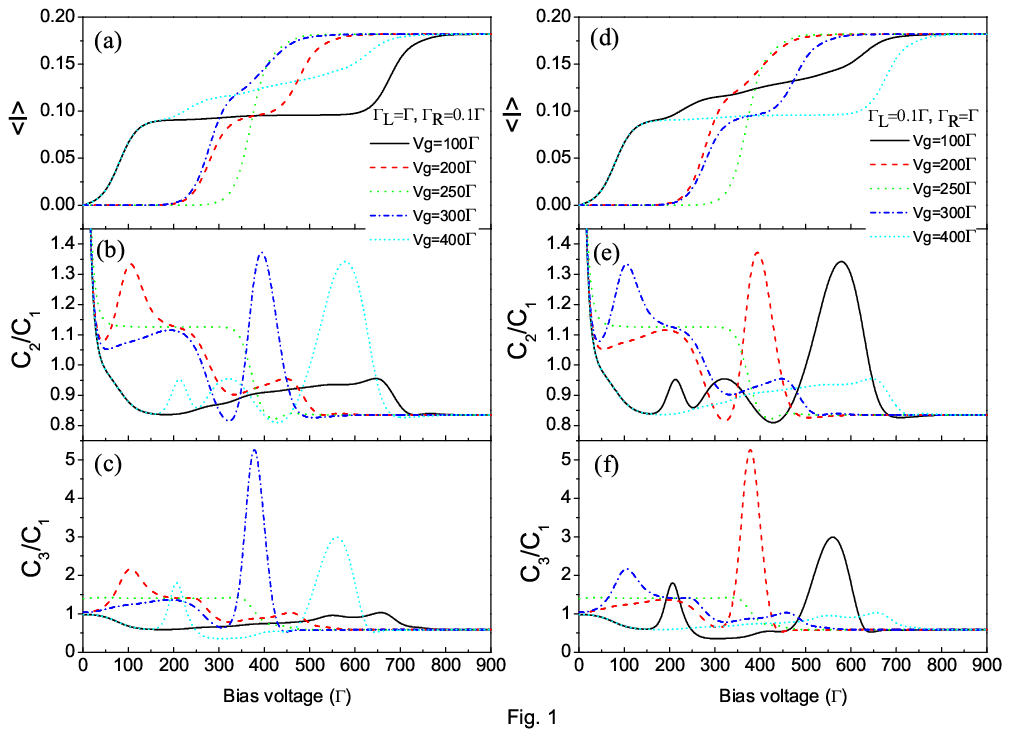}
\end{center}
\caption{(Colour online) The first three cumulants of zero-frequency
current fluctuation versus bias voltage for different gate voltages.
(a), (b) and (c) for $\Gamma_{L}/\Gamma_{R}=10$, (d), (e) and (f)
for $\Gamma_{L}/\Gamma_{R}=0.1$. The molecular parameters: $S=2$,
$\varepsilon_{d}=200\Gamma$, $J=100\Gamma$, $K_{2}=40\Gamma$,
$B=80\Gamma$ and $k_{B}T=10\Gamma$.}%
\end{figure}

\begin{figure}[t]
\begin{center}
\includegraphics[height=10cm,width=12cm]{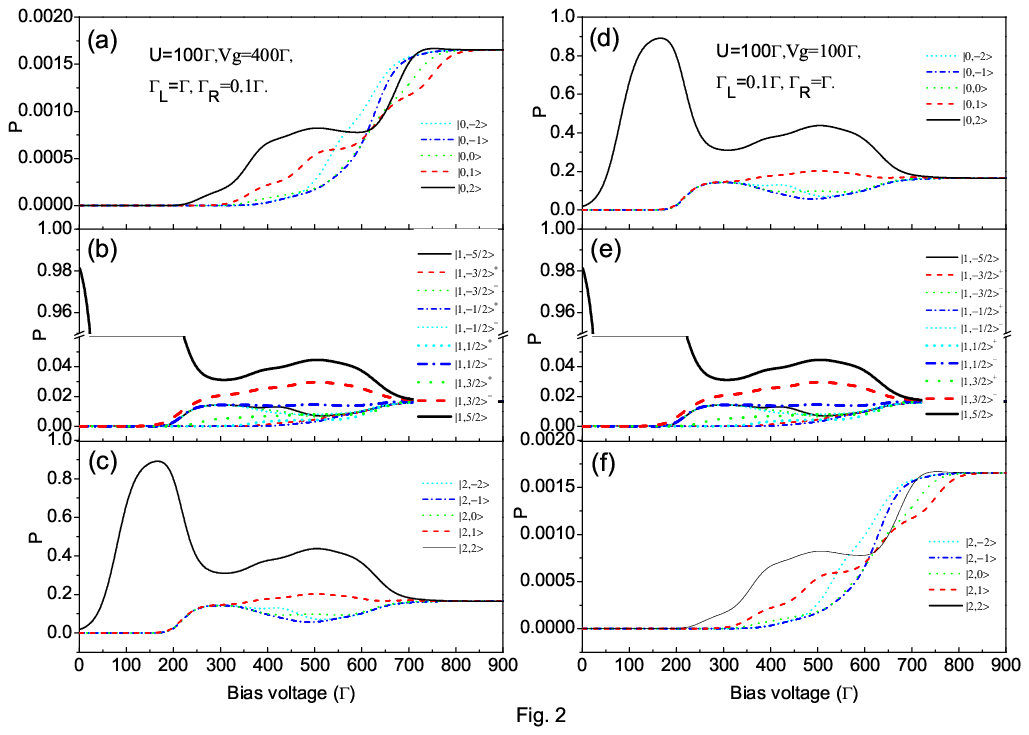}
\end{center}
\caption{(Colour online) The probability distribution of molecular
eigenstates versus bias voltage for different gate voltages with
$U=100\Gamma$. (a), (b) and (c) $V_{g}=400\Gamma$ and $\Gamma_{L}/\Gamma_{R}=10$, (d), (e) and (f) $%
V_{g}=100\Gamma$ and $\Gamma_{L}/\Gamma_{R}=0.1$. The molecular
parameters are the same as in Fig. 1.}%
\end{figure}

\begin{figure}[t]
\begin{center}
\includegraphics[height=8cm,width=10cm]{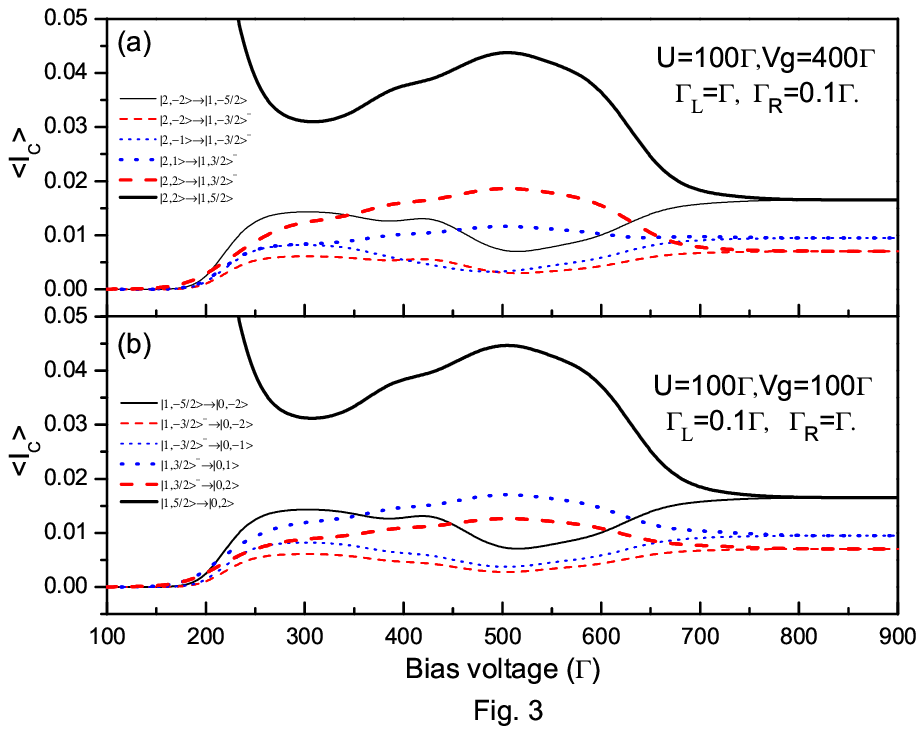}
\end{center}
\caption{(Colour online) The channel currents versus bias voltage
for various gate voltages with $U=100\Gamma$. (a) $V_{g}=400\Gamma$ and $%
\Gamma_{L}/\Gamma_{R}=10$, (b) $V_{g}=100\Gamma$ and $\Gamma_{L}/%
\Gamma_{R}=0.1$. The $\left\vert i\right\rangle
\rightarrow\left\vert j\right\rangle $ denotes the molecular channel
current $I_{\left\vert i\right\rangle \rightarrow\left\vert
j\right\rangle }$. The molecular parameters are the same as in Fig.
1.}%
\end{figure}

\begin{figure}[t]
\begin{center}
\includegraphics[height=10cm,width=12cm]{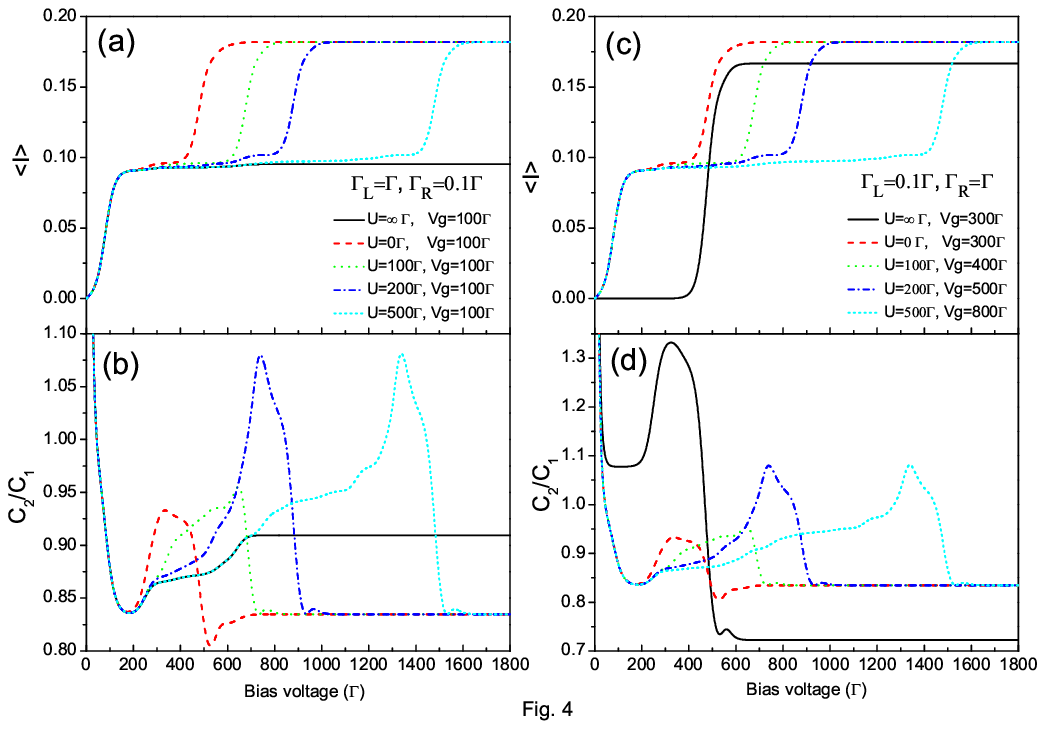}
\end{center}
\caption{(Colour online) The average current and shot noise versus
bias voltage for the different Coulomb interaction energies. (a) and (b) for $%
\Gamma _{L}/\Gamma_{R}=10$, (c) and (d) for
$\Gamma_{L}/\Gamma_{R}=0.1$. The molecular parameters are the same
as in Fig. 1.}%
\end{figure}

\begin{figure}[t]
\begin{center}
\includegraphics[height=10cm,width=12cm]{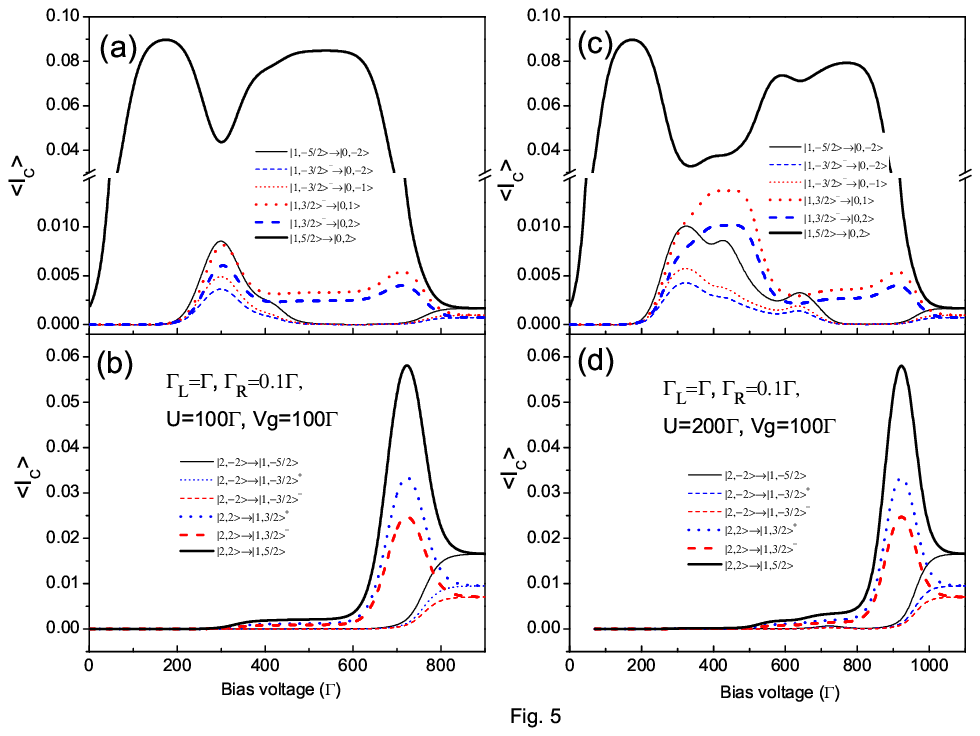}
\end{center}
\caption{(Colour online) The channel currents versus bias voltage
for different Coulomb interaction energies with
$\Gamma_{L}/\Gamma_{R}=10$. (a) and (b) $U=100\Gamma$,
$V_{g}=100\Gamma$,
 (c) and (d) $U=200\Gamma$, $V_{g}=100\Gamma $. The channel current expression and molecular
parameters are the same as Fig. 3.}%
\end{figure}

\begin{figure}[t]
\begin{center}
\includegraphics[height=10cm,width=12cm]{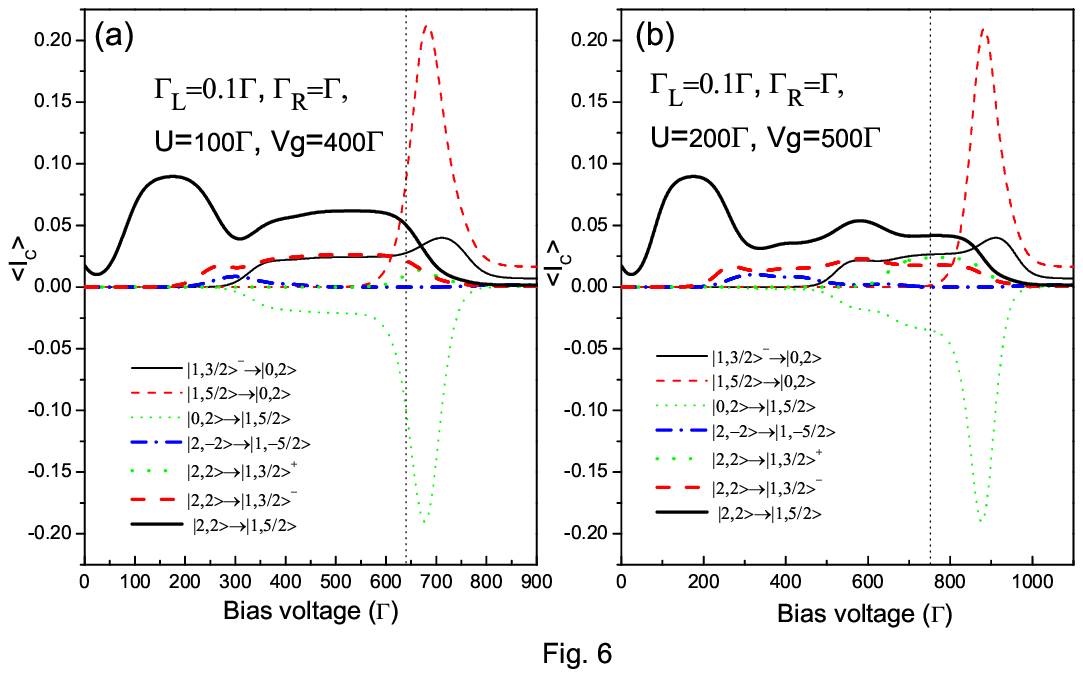}
\end{center}
\caption{(Colour online) The channel currents versus bias voltage
for different Coulomb interaction energies with
$\Gamma_{L}/\Gamma_{R}=0.1$. (a) for $U=100\Gamma$,
$V_{g}=400\Gamma$,
 (b) $U=200\Gamma$, $V_{g}=500\Gamma $. The channel current expression and molecular
parameters are the same as Fig. 3.}%
\end{figure}

\begin{figure}[t]
\begin{center}
\includegraphics[height=8cm,width=10cm]{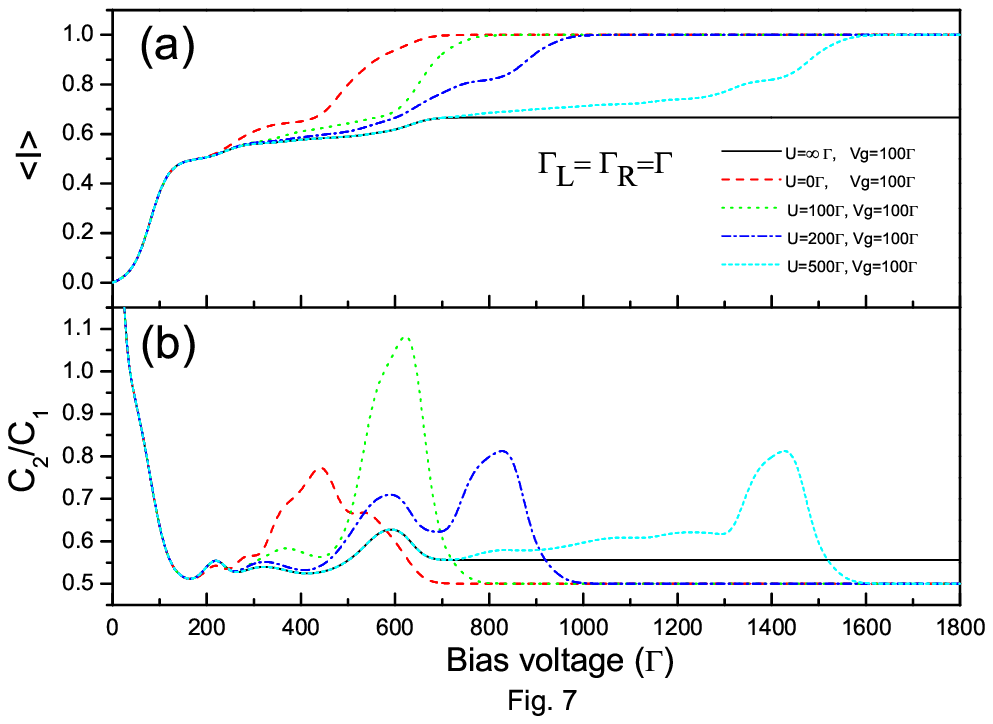}
\end{center}
\caption{(Colour online) The average current and shot noise versus
bias voltage for different Coulomb interaction energies with
$\Gamma_{L}=\Gamma _{R}=\Gamma$. The molecular parameters are the
same as in Fig. 1.}%
\end{figure}

\end{document}